\documentclass[twocolumn,aps,prb,superscriptaddress]{revtex4}

\usepackage{graphicx}
\usepackage{dcolumn}
\usepackage{bm}
\usepackage[cmex10]{amsmath}
\usepackage{sidecap}

\usepackage{polski}
\usepackage[T1]{fontenc}
\usepackage{array,multirow}

\begin{document}

\preprint{APS/123-QED}

\title{Understanding stability diagram of perpendicular magnetic tunnel junctions}

\author{Witold Skowro\'{n}ski}
 \email{skowron@agh.edu.pl}
\author{Maciej Czapkiewicz}
 \email{czapkiew@agh.edu.pl}
\author{S\l{}awomir Zi\k{e}tek}
\author{Jakub Ch\k{e}ci\'{n}ski}
\author{Marek Frankowski}
\author{Piotr Rzeszut}
\affiliation{AGH University of Science and Technology, Department of Electronics, Al. Mickiewicza 30, 30-059 Krak\'{o}w, Poland}
\author{Jerzy Wrona}
\affiliation{Singulus Technologies, Kahl am Main, 63796, Germany}

\date{\today}

\begin{abstract}
Perpendicular magnetic tunnel junctions (MTJ) with a bottom pinned reference layer and a composite free layer (FL) are investigated. Different thicknesses of the FL were tested to obtain an optimal balance between tunneling magnetoresistance (TMR) ratio and perpendicular magnetic anisotropy. After annealing at 400 $^\circ$C, the TMR ratio for 1.5 nm thick CoFeB sublayer reached 180 \% at room temperature and 280 \% at 20 K with an MgO tunnel barrier thickness corresponding to the resistance area product RA = 10 Ohm$\mathrm{\mu}$m$^2$. The voltage vs. magnetic field stability diagrams measured in pillar-shaped MTJs with 130 nm diameter indicate the competition between spin transfer torque (STT), voltage controlled magnetic anisotropy (VCMA) and temperature effects in the switching process. An extended stability phase diagram model that takes into account all three effects and the effective damping measured independently using broadband ferromagnetic resonance technique enabled the determination of both STT and VCMA coefficients that are responsible for the FL magnetization switching. 

\end{abstract}

\maketitle

\section{Introduction}
Magnetic tunnel junctions (MTJs) have become a basic building block for various types of spintronics devices, such as magnetic random access memory (MRAM) cells, magnetic field sensors and microwave generators or detectors \cite{stamps_2014_2014}. The properties of spintronics devices, such as thermal stability of an MRAM cell \cite{ikeda_perpendicular-anisotropy_2010}, or sensitivity of microwave detectors \cite{fang_giant_2016} utilizin  g MTJs can be greatly improved by using magnetic layers with perpendicular anisotropy \cite{yakata_influence_2009}. Among a few ways to realize such a perpendicular MTJ, taking advantage of the interface anisotropy component \cite{koziol-rachwal_room-temperature_2013} yields the best results so far, especially in terms of high tunneling magnetoresistance ratio (TMR), which is measured typically in MTJs with CoFeB/MgO/CoFeB trilayer. Recent studies on perpendicular MTJ showed the TMR ratio exceeding 200\% \cite{tezuka_perpendicular_2016} thanks to careful optimization of both the free layer (FL) and reference layer (RL) structure \cite{yakushiji_perpendicular_2016}. In addition, one of the key challenges for the commercial development of spin transfer torque (STT)-MRAM is to optimize perpendicular MTJ to withstand the temperature budget introduced at the back end of line CMOS fabrication process with temperatures up to 400 $^\circ$C. To achieve this a careful design of the layer stack, taking into account all constituent layers as well as the properties and the treatment of the bottom electrode, has to be performed.

In this letter, we report on the perpendicular MTJ with a composite CoFeB/W/CoFeB FL \cite{yakushiji_ultralow-voltage_2013, sato_properties_2014}, which is characterized by high perpendicular magnetic anisotropy and spin polarization resulting in up to 180 \% TMR measured at room temperature and above 280 \% TMR at low temperature. The RL is pinned to a synthetic ferromagnet (SyF) consisting of Co/Pt super-latices \cite{kugler_Co-Pt_2012} coupled by a thin Ru spacer. Electrical transport measurements were performed in MTJs patterned into 130-nm diameter pillar. Voltage vs. perpendicular magnetic field switching diagrams \cite{worledge_spin_2011, oh_bias-voltage_2009} are measured in order to separate between STT, voltage control of magnetic anisotropy (VCMA) and temperature effects. An analytic model based on work by Bernert et al. \cite{bernert_phase_2014} was extended to reproduce the experimental results.

\section{Experiment}
The multilayers with the following structure were deposited: buffer / SyF / separator /CoFeB(1) / MgO(0.82) / CoFeB($t_\mathrm{FL}$) / W(0.3) / CoFeB(0.5) / MgO (0.76) / capping  (thicknesses in nm), with $t_\mathrm{FL}$ ranging from 1 up to 1.6 nm, using Singulus TIMARIS sputtering system. The bottom Co/Pt super-lattices coupled by a thin Ru spacer are characterized by high perpendicular magnetic anisotropy (PMA). The Ta/Co/W-based separator ensures high ferromagnetic coupling between the top super-lattice and the RL. In addition, it provides structural transition from a face center cubic SyF \cite{kanak_influence_2007} to a body center cubic CoFeB and contributes to the absorption of B atoms from CoFeB during annealing and crystallization processes.

After the deposition, the samples were annealed at 400 $^\circ$C to induce proper crystallographic orientation of Fe-rich CoFeB and PMA of the CoFeB/MgO interfaces. Wafer-level parameters of the deposited multilayers were investigated by current in-plane tunneling (CIPT) \cite{worledge_magnetoresistance_2003}, vibrating sample magnetometry (VSM) and broadband ferromagnetic resonance (FMR) methods \cite{silva_inductive_1999}. The latter was performed by measuring the complex transmission coefficient (S$_\mathrm{21}$) in a dedicated coplanar waveguide with a 10 $\times$ 8 mm unpatterned sample placed face down. The frequency of the vector network analyzer is kept between 4 and 16 GHz, while sweeping the perpendicular magnetic field in $\pm$ 550 kA/m range. 

After the above mentioned wafer-level measurements, selected MTJs were patterned into circular cross-section pillars with diameter ranging from 130 up to 980 nm by means of electron-beam lithography, ion-beam etching and lift-off process.  

The transport properties presented in this work were measured for the smallest devices with the area of $A$ = 0.013 $\mathrm{\mu} m^2$ in a dedicated probe station equipped with magnetic field source. Four-probe method with a voltage source was used to apply 1-ms long pulses and measure the resistance during this voltage-pulse application. The stability diagrams were determined by sweeping the voltage pulses amplitude in the presence of a given magnetic field. Selected devices were characterized at low temperatures of $T$ = 20 K in order to determine the temperature influence on the magnetization switching properties. 

\section{Modelling}
Magnetization direction of the FL ($\vec{m}_\mathrm{FL}$) was calculated based on the Landau-Lifschitz-Gilbert (LLG) equation with the following STT components taken into account:
\begin{eqnarray}
	\label{eq:llg_stt}
	\frac{d \vec{m}_{\mathrm{FL}}}{dt} = -\gamma_0 \vec{m}_{\mathrm{FL}} \times \vec{H_{\mathrm{eff}}} + \alpha \vec{m}_{\mathrm{FL}} \times \frac{d\vec{m}_{\mathrm{FL}}}{dt} \nonumber \\ - \gamma_0 a_{\parallel} \frac{VR_\mathrm{P}}{R} \left(\vec{m}_{\mathrm{FL}} \times \left( \vec{m}_{\mathrm{FL}} \times \vec{m}_{\mathrm{RL}} \right) \right) \nonumber \\ - \gamma_0 a_{\perp} \left(\frac{VR_\mathrm{P}}{R}\right)^2 \left( \vec{m}_{\mathrm{FL}} \times \vec{m}_{\mathrm{RL}} \right)
\end{eqnarray}
where $\gamma_0$ = $\gamma$ $\mu_0$, with the gyromagnetic ratio $\gamma$ = ($g \mu_\mathrm{B}$)/$\hbar$ = 28 GHz/T,  $\mu_0$ is the permeability of the free space, $g$ is the Lande spectroscopic splitting factor, $\mu_\mathrm{B}$ is the Bohr magneton, $\hbar$ is the reduced Planck's constant, $a_\parallel$ and $a_\perp$ are the parallel and perpendicular STT coefficients expressed in T/V  and T/V$^2$ units, respectively, $\alpha$ is the magnetization damping, $R$ and $R_\mathrm{P}$ are the MTJ resistance in a given state and minimal (parallel state) resistance, $H_\mathrm{eff}$ is the effective magnetic field: $H_\mathrm{eff}$ = $H$ $\pm$ $H_\mathrm{W}$ + $H_\mathrm{S}$, where, $H$ is the external perpendicular field, $H_\mathrm{W}$ is the switching field and $H_\mathrm{S}$ is the offset field.

Stability diagram was modeled based on Ref.\cite{bernert_phase_2014}:
\begin{equation}
V_\mathrm{C} = \frac{a_{\parallel}R}{2\alpha a_{\perp}R_\mathrm{P}} - \sqrt{\left(\frac{a_{\parallel}R}{2\alpha a_{\perp}R_\mathrm{P}}\right)^2-\frac{\mu_0 R^2}{a_\perp R_\mathrm{P}^2} H_\mathrm{eff}}
\label{eq:bernert}
\end{equation}
where, $V_\mathrm{C}$ is the switching voltage. It was assumed that parallel (perpendicular) torque component is a linear (quadratic) function of the applied current \cite{skowronski_influence_2013}. To account for the additional physical effects that contribute to the stability diagram, namely VCMA and temperature effects, $H_\mathrm{W}$ is scaled by the factor: 
\begin{equation}
H_\mathrm{W} = H_\mathrm{C}\left( 1-k_\mathrm{V}V- \sqrt{k_\mathrm{t} T} \right)
\label{eq:correction}
\end{equation}
where $V$ is the applied voltage, $H_\mathrm{C}$ is the coercive field, $k_\mathrm{V}$ is the VCMA coefficient \cite{shiota_quantitative_2011, skowronski_magnetic_2012} and $T$ is the ambient temperature. The dependence of the switching field on the temperature is represented by $k_\mathrm{t}$, which in the first approximation is a square-root function \cite{raquet_dynamical_1995}. 

\begin{figure}
\centering
\includegraphics[width=\columnwidth]{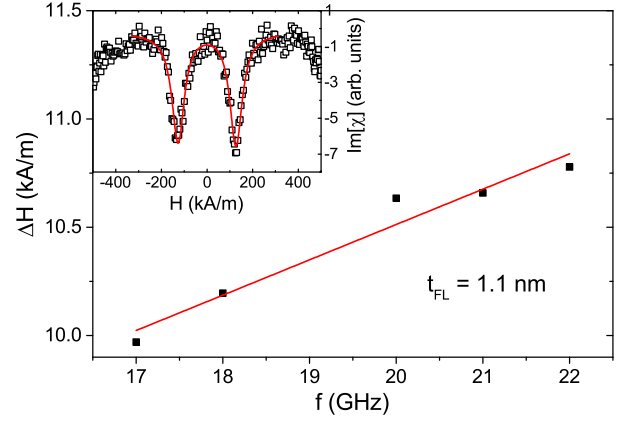}
\caption{FMR linewidth (full symbols) as a function of the excitation frequency for MTJ with $t_\mathrm{FL}$ = 1.1 nm together with a fit (solid line) to the Eq. \ref{eq:delta_H}. Inset presents the measured imaginary part of the magnetic susceptibility (open symbols) as a function of the magnetic field for $f$ = 19 GHz together with a fit to the Eq. \ref{eq:susceptibility} (solid line).}
\label{fig:fmr}
\end{figure}

The damping factor was measured independently by the broadband FMR technique. For each microwave frequency $f$, the complex magnetic susceptibility vs. magnetic field $\chi(H)$ is extracted from S$_\mathrm{21}$ measurement by subtracting the magnetic independent offset and time-dependent drift \cite{nembach_perpendicular_2011}: 

\begin{equation}
\chi(H) = \frac{M_\mathrm{eff}(H-M_\mathrm{eff})}{(H-M_\mathrm{eff})^2 - H_\mathrm{f}^2 - i \frac{\Delta H}{2}(H-M_\mathrm{eff})}
\label{eq:susceptibility}
\end{equation}

where $M_\mathrm{eff}$ = $M_\mathrm{S}$ - $H_\mathrm{K}$ is the effective magnetization, magnetization saturation and perpendicular magnetic anisotropy field, respectively, $\Delta H$ is the linewidth and $H_\mathrm{f}$ = $2\pi f$ / $(\gamma \mu_0)$. 
Figure \ref{fig:fmr} presents the dependence of the $\Delta H$ on the excitation frequency fitted by the Eq. \ref{eq:delta_H}:
\begin{equation}
\Delta H = \frac{4 \pi \alpha f}{\gamma_0} + \Delta H_0
\label{eq:delta_H}
\end{equation}

\section{Results and discussion}
\begin{figure}
\centering
\includegraphics[width=\columnwidth]{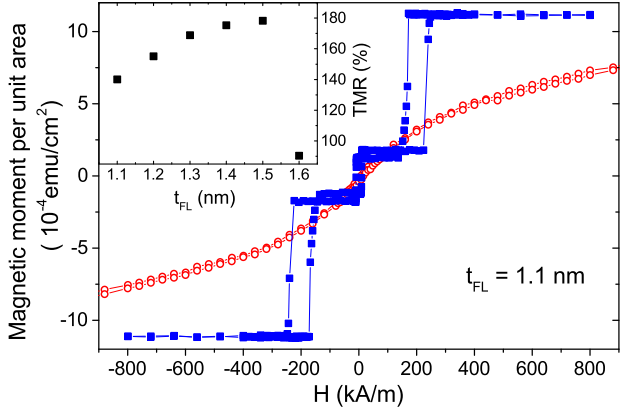}
\caption{Magnetic moment per unit area vs. in-plane (open symbols) and perpendicular (full symbols) magnetic field of the MTJ with $t_\mathrm{FL}$ = 1.1 nm. Inset presents the TMR ratio dependence on $t_\mathrm{FL}$ measured using CIPT method.}
\label{fig:cipt}
\end{figure}

VSM measurements of a representative sample with $t_\mathrm{FL}$ = 1.1 nm presented in Fig. \ref{fig:cipt} reveals independent switching of the FL (at small magnetic fields below 50 kA/m) and RL (at high magnetic fields between 150 and 300 kA/m), which ensures bistable parallel (P) and antiparallel (AP) state. The FL magnetization was calculated and yielded $\mu_0 M_\mathrm{S}$ = 1.12 T. An inset of Fig. \ref{fig:cipt} depicts the TMR ratio for different $t_\mathrm{FL}$ measured on the wafer-level using CIPT method. An increase of the TMR from 140\% for $t_\mathrm{FL}$ = 1.1 nm up to TMR = 180\% for $t_\mathrm{FL}$ = 1.5 nm is explained by an increase of the spin polarization for the thicker ferromagnetic layer. A rapid reduction of the TMR for $t_\mathrm{FL}$ = 1.6 nm is caused by the transition of the FL magnetization vector to the sample plane. After the patterning process the TMR ratio dropped by about 5-15\%, which is explained by the appearance of the small serial parasitic resistance that has negligible influence on the MTJ parameters derived afterwards.

\begin{figure}
\centering
\includegraphics[width=\columnwidth]{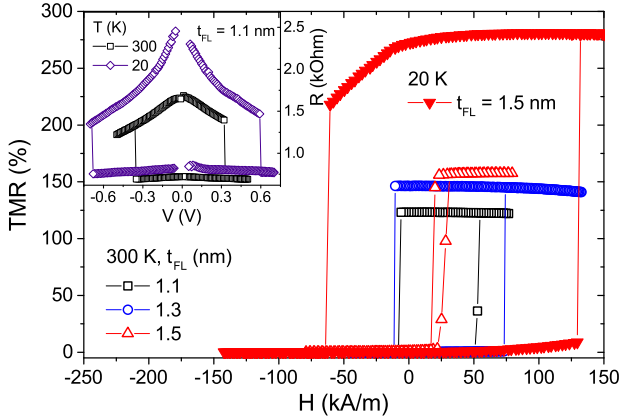}
\caption{TMR vs. magnetic field loop of MTJs with different $t_\mathrm{FL}$ measured at room temperature (300 K). Significantly smaller coercive field $H_\mathrm{C}$ is measured for $t_\mathrm{FL}$ = 1.5 nm, which increases at T = 20 K. An inset presents a resistance vs. voltage switching loop measured in an external magnetic field of $H$ = 25 kA/m, obtained at $T$ = 300 K (squares) and $T$ = 20 K (circles).}
\label{fig:tmr}
\end{figure}

Figure \ref{fig:tmr} presents the TMR vs. magnetic field dependence measured in the MTJ with different $t_\mathrm{FL}$. Increase in $t_\mathrm{FL}$ leads to an increase in TMR ratio and decrease in the coercive field. The offset field of about $H_\mathrm{s}$ = 25 kA/m originates from the stray field, which depends on the MTJ lateral size. 

\begin{figure}
\centering
\includegraphics[width=\columnwidth]{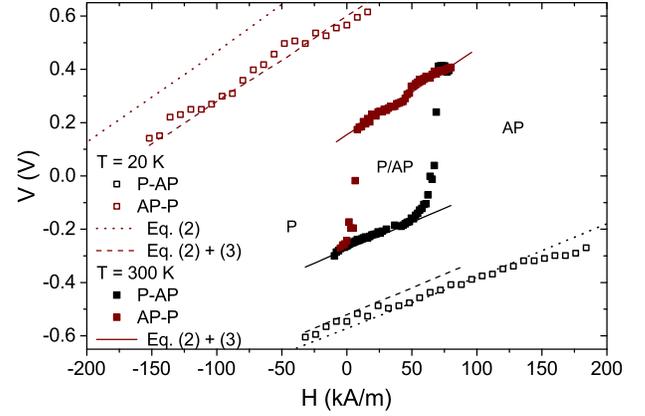}
\caption{Voltage vs. magnetic field stability diagram measured in the MTJ with $t_\mathrm{FL}$ = 1.1 nm at T = 20 K (open symbols) and T = 300 K (full symbols). Dotted lines represents approximation based on Eq. \ref{eq:bernert}. An extended model based on correction presented in Eq. \ref{eq:correction} is represented by dashed (20 K) and solid (300 K) lines.}
\label{fig:cims}
\end{figure}

To further elucidate on the properties of the fabricated MTJ, current (voltage)-induced switching loops were measured in a presence of the perpendicular magnetic field. Inset of Fig. \ref{fig:tmr} presents a representative resistance vs. voltage loop measured in a magnetic field compensating the offset field in the MTJ with $t_\mathrm{FL}$ = 1.1 nm. It has been already established that apart from the conventional STT effect observed in MTJs with relatively thin MgO barriers, the switching process can be also affected by the VCMA effect in a thin FL \cite{zhu_voltage-induced_2012, shiota_induction_2011}. To investigate the switching process in more detail, we repeated the R(V) loop measurements, with different constant magnetic field. The stability diagram obtained in this way both at room temperature ($T$ = 300 K) and low temperature ($T$ = 20 K) for the MTJ with $t_\mathrm{FL}$ = 1.1 nm is presented in Fig. \ref{fig:cims}. 
To understand these diagrams, the following fitting procedure was used. First, low-temperature data were modeled using Eq. \ref{eq:bernert} to obtain $H_\mathrm{W}$, $H_\mathrm{S}$ (being the offset field measured at low bias voltage) and STT coefficients, which are little affected by heat. The slope of V(H) depends mostly on $a_\parallel$, whereas, the vertical offset is adjusted by $H_\mathrm{W}$ - solid lines in Fig. \ref{fig:cims}. Next, to compensate the offset between AP-P and P-AP switching voltages (which take place at opposite electric field applied to the MgO/CoFeB interface) $k_\mathrm{V}$ was introduced according to the Eq. \ref{eq:correction}, without temperature influence so far ($k_\mathrm{t}$ = 0). Finally, thermal reduction of $H_\mathrm{W}$ was introduced by adjusting $k_\mathrm{t}$ to fit the stability diagram obtained at room temperature. 
In addition, for the precise derivation of the STT coefficients, the magnetization damping was calculated based on an independent FMR measurement presented above and included in Tab. \ref{tab:stt}. 

Fitting the experimental stability diagram to the Eq. \ref{eq:bernert} and \ref{eq:correction} yielded the temperature coefficients of $k_\mathrm{t}$ = 0.0014 1/K. This parameter was kept constant for MTJs with different $t_\mathrm{FL}$. Remaining parameters of the stability diagrams for each $t_\mathrm{FL}$ were modeled independently. 
For $t_\mathrm{FL}$ = 1.1 nm, the following STT components were obtained $a_\parallel$ = 0.024 T/V and $a_\perp$ = 0.02 T/V$^2$, however, we note that the modeled stability diagram is only little sensitive to $a_\perp$, which agrees with another macrospin approach based on LLG equation presented in Ref. \cite{timopheev_respective_2015}. In-plane torque $\tau_\parallel$ was thereafter recalculated using Eq. \ref{eq:torque}:

\begin{equation}
\tau_\parallel = \frac{M_\mathrm{S} \upsilon \tau_\mathrm{LLG}}{\gamma}
\label{eq:torque}
\end{equation}
where $\upsilon$ is the FL volume and $\tau_\mathrm{LLG}$ = -$\left( \gamma a_\parallel \right) $. As the result we obtained $\tau_\parallel$ = 4.5$\times$10$^{-19}$ Nm/V, which agrees well with literature values of STT in case of an in-plane MTJ \cite{sankey_measurement_2007, kubota_quantitative_2007, skowronski_influence_2013}. 
Regarding the VCMA, the best results for MTJ with $t_\mathrm{FL}$ = 1.1 nm were obtained for $k_\mathrm{V}$ = 0.12 1/V. Based on the following relation: $k_\mathrm{V}$ = $\kappa$/$\mu_0 H_\mathrm{C} M_\mathrm{S} t_\mathrm{FL} t_\mathrm{B}$, where $t_\mathrm{B}$ = 0.82 nm is the tunnel barrier thickness, VCMA coefficient of $\kappa$ = 46 fJ/Vm was calculated, which fits well the commonly measured values for CoFeB/MgO devices \cite{skowronski_underlayer_2015, alzate_temperature_2014}.VCMA and STT coefficients of all investigated MTJs are gathered in Tab. \ref{tab:stt}.




\begin{table}[!t]
\centering
\caption{Summary of the obtained perpendicular MTJ parameters.}
\label{tab:stt}
\begin{tabular}{llllll}
\hline
\multicolumn{1}{c}{$t_\mathrm{FL}$} & \multicolumn{1}{c}{$H_\mathrm{C}$} & \multicolumn{1}{c}{$\kappa$} & \multicolumn{1}{c}{$\tau_\parallel$} & \multicolumn{1}{c}{$\alpha$} & \multicolumn{1}{c}{$M_\mathrm{eff}$} \\
\multicolumn{1}{c}{(nm)} & \multicolumn{1}{c}{(kA/m)} & \multicolumn{1}{c}{(fJ/Vm)} & \multicolumn{1}{c}{(Nm)} & \multicolumn{1}{c}{no units} & \multicolumn{1}{c}{(kA/m)} \\
\hline
\multicolumn{1}{c}{1.1} & \multicolumn{1}{c}{264} & \multicolumn{1}{c}{46} &  \multicolumn{1}{c}{4.5$\times 10^{-19}$} & \multicolumn{1}{c}{0.038} & \multicolumn{1}{c}{-450}\\ 
\multicolumn{1}{c}{1.3} & \multicolumn{1}{c}{280} & \multicolumn{1}{c}{73} &  \multicolumn{1}{c}{4.5$\times 10^{-19}$} & \multicolumn{1}{c}{0.044} & \multicolumn{1}{c}{-15}\\
\multicolumn{1}{c}{1.5} & \multicolumn{1}{c}{72} & \multicolumn{1}{c}{66} & \multicolumn{1}{c}{5.9$\times 10^{-19}$} & \multicolumn{1}{c}{0.087} & \multicolumn{1}{c}{-0.5}\\
\hline
\end{tabular}
\end{table} 

The in-plane torque component obtained from the stability diagram is almost constant as a function of $t_\mathrm{FL}$, which is explained by little dependence of the TMR ratio, and thus the spin polarization on the ferromagnetic layer thickness in the investigated regime. The VCMA coefficient is comparable for MTJs with $t_\mathrm{FL}$ = 1.3 nm and 1.5 nm and greater than in MTJ with $t_\mathrm{FL}$ = 1.1 nm. This behavior is expected, as for thicker $t_\mathrm{FL}$ the absolute value of the effective magnetization is reduced and it is more susceptible to the anisotropy change induced by the electric field \cite{nozaki_large_2016}. Moreover, in the same thickness regime, where the transition between perpendicular and in-plane magnetization occurs, the effective damping increases, which may be attributed to an increase in the level of magnetization disorder \cite{frankowski_perpendicular_2017}.

\section{Summary}
In summary, we investigated perpendicular MTJs with composite CoFeB/W/CoFeB FL of different thickness and SyF Co/Pt/Ru-pinned RL. In the investigated FL thickness range we observed an increase of the effective damping extracted from the broadband FMR measurements with increasing FL thickness, which is mainly caused by the reduction of the effective magnetization. After patterning MTJs into nano-meter scale pillars, we measured the resistance vs. voltage loops and created the stability diagrams for each FL thickness. To model the experimental data, we included the thermal and VCMA terms into the theoretical STT-switching phase diagram. Based on the fitting procedure, we obtained STT components together with the VCMA coefficient. Our findings shine more light on the switching process of MTJs applied in future MRAM technologies.


\section*{Acknowledgments}
The authors would like to express their gratitude to Prof. T. Stobiecki for a fruitful discussion and his critical remarks. The project is supported by Polish National Center for Research and Development grant No. LIDER/467/L-6/14/NCBR/2015. Nanofabrication process was performed at Academic Center for Materials and Nanotechnology of AGH University. J.Ch. acknowledges the scholarship under Marian Smoluchowski Krakow Research Consortium KNOW programme. Numerical calculations were supported by PL-GRID infrastructure.

\bibliographystyle{unsrt}
\bibliography{Skowronski_library}


\end{document}